\documentclass[letterpaper]{jpconf}
\usepackage{graphicx}
\usepackage{amsmath}
\begin{document}
\title{\textit{B} Physics at CDF}

\author{Elisa Pueschel (on behalf of CDF Collaboration)}

\address{Carnegie Mellon University}

\begin{abstract}
We present the latest $B$ physics results from the CDF experiment at the Fermilab Tevatron collider. We focus on a number of analyses, including a measurement of the forward-backward asymmetry of $B\rightarrow K^{(*)}\mu\mu$ decays, determination of the \textit{CP} violating phase $\sin2\beta_{s}$ in $B^{0}_{s}\rightarrow J/\psi \phi$ decays, $B \rightarrow J/\psi X$ lifetime measurements, observation of resonance structure in $\Lambda_{b} \rightarrow \Lambda_{c} \pi^{-}\pi^{+}\pi^{-}$, and $\Upsilon(1S)$ polarization.
\end{abstract}

\section{Introduction}

The Tevatron is a $p\bar{p}$ collider, with collisions occuring at 1.96 TeV center of mass energy. As a hadron collider, the Tevatron has access to all bottom species, including $B^{0}$, $B^{+}$, $B^{0}_{s}$, $B^{+}_{c}$ and $\Lambda^{0}_{b}$ hadrons. The hadronic detector environment is complex, with large amounts of background. The CDF experiment employs sophisticated triggers to select $B$ decays. Additionally, CDF's precise momentum and vertexing resolution facilitate a variety of $B$ physics measurements, ranging from lifetime and $CP$ violation studies to $B$ hadron spectroscopy measurements.

\section{Forward-backward asymmetry in $\boldsymbol{B\rightarrow K^{(*)}\mu\mu}$}

Flavor changing neutral current processes can occur via penguin diagrams in the standard model.  The transition $b \rightarrow s\ell\ell$, for instance, is a FCNC process, present in the decays $B^{+}\rightarrow K^{+}\mu^{+}\mu^{-}$, $B^{0}\rightarrow K^{*0}\mu^{+}\mu^{-}$ and $B_{s}\rightarrow \phi\mu^{+}\mu^{-}$. The rates for these decays could be enhanced by new physics contributions to the penguin diagrams. This would consequently alter the differential branching ratio and forward-backward asymmetry for these decays from the standard model predictions.

$B^{+}\rightarrow K^{+}\mu^{+}\mu^{-}$, $B^{0}\rightarrow K^{*0}\mu^{+}\mu^{-}$ and $B_{s}\rightarrow \phi\mu^{+}\mu^{-}$ decays are reconstructed using 4.4 fb$^{-1}$ of data from a di-muon trigger. The observed signal yields for the three decay modes are 120$\pm$16, 101$\pm$12, and 27$\pm$6 events, respectively. This measurement is the first observation of $B_{s}\rightarrow \phi\mu^{+}\mu^{-}$ decays at a 6.3$\sigma$ significance. The absolute branching ratios of the three modes are measured to be: $BR(B^{+}\rightarrow K^{+}\mu^{+}\mu^{-})$ = 0.38$\pm$0.05(stat)$\pm$0.03(syst)$\times$10$^{-6}$, $BR(B^{0}\rightarrow K^{*0}\mu^{+}\mu^{-})$ = 1.06$\pm$0.14(stat)$\pm$0.09(syst)$\times$10$^{-6}$, $BR(B^{0}_{s}\rightarrow \phi\mu^{+}\mu^{-})$ = 1.44$\pm$0.33(stat)$\pm$0.46(syst)$\times$10$^{-6}$.

\begin{figure}[htbp]
\centerline{
\makebox{\includegraphics[width=0.35\textwidth]{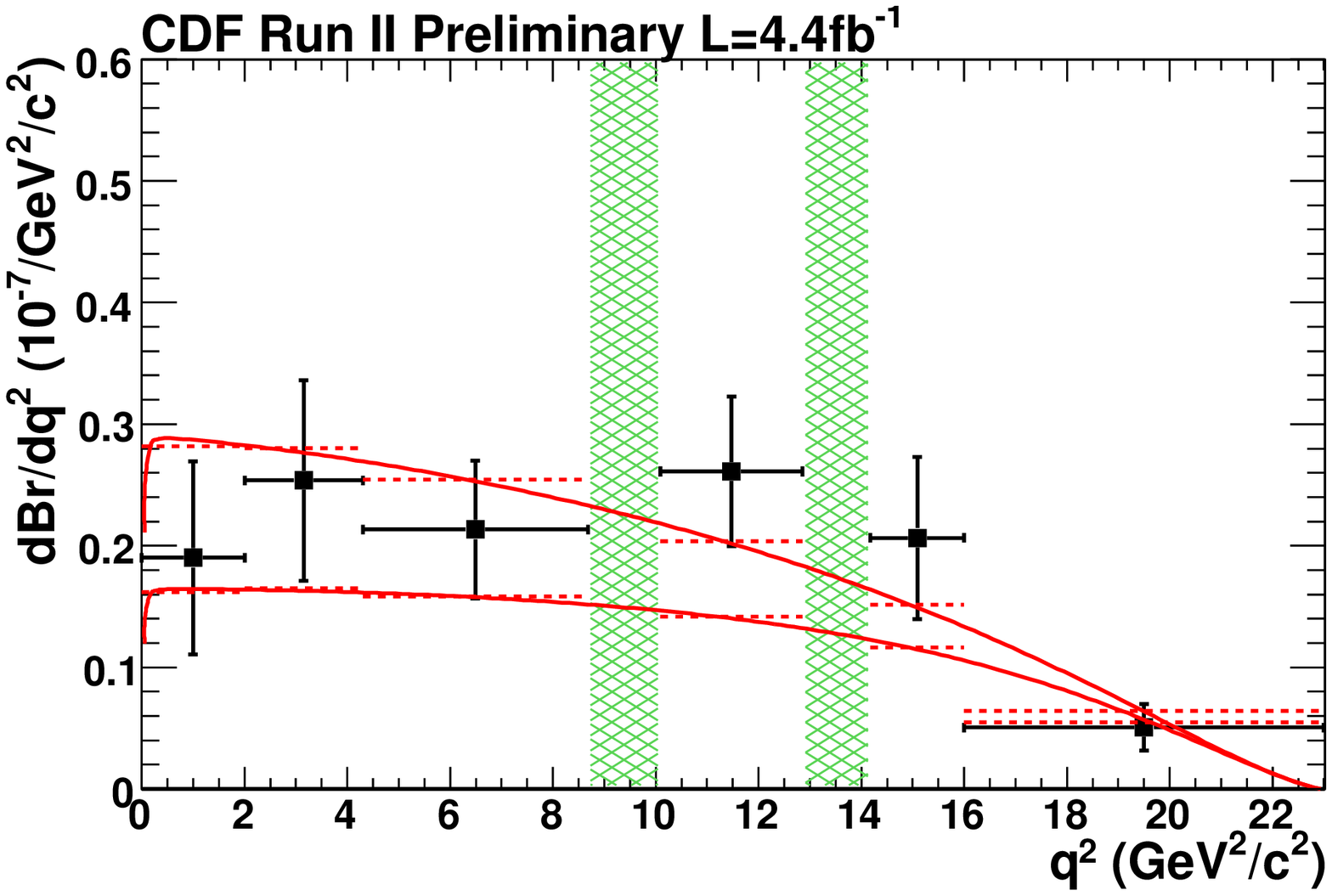}}
\makebox{\includegraphics[width=0.35\textwidth]{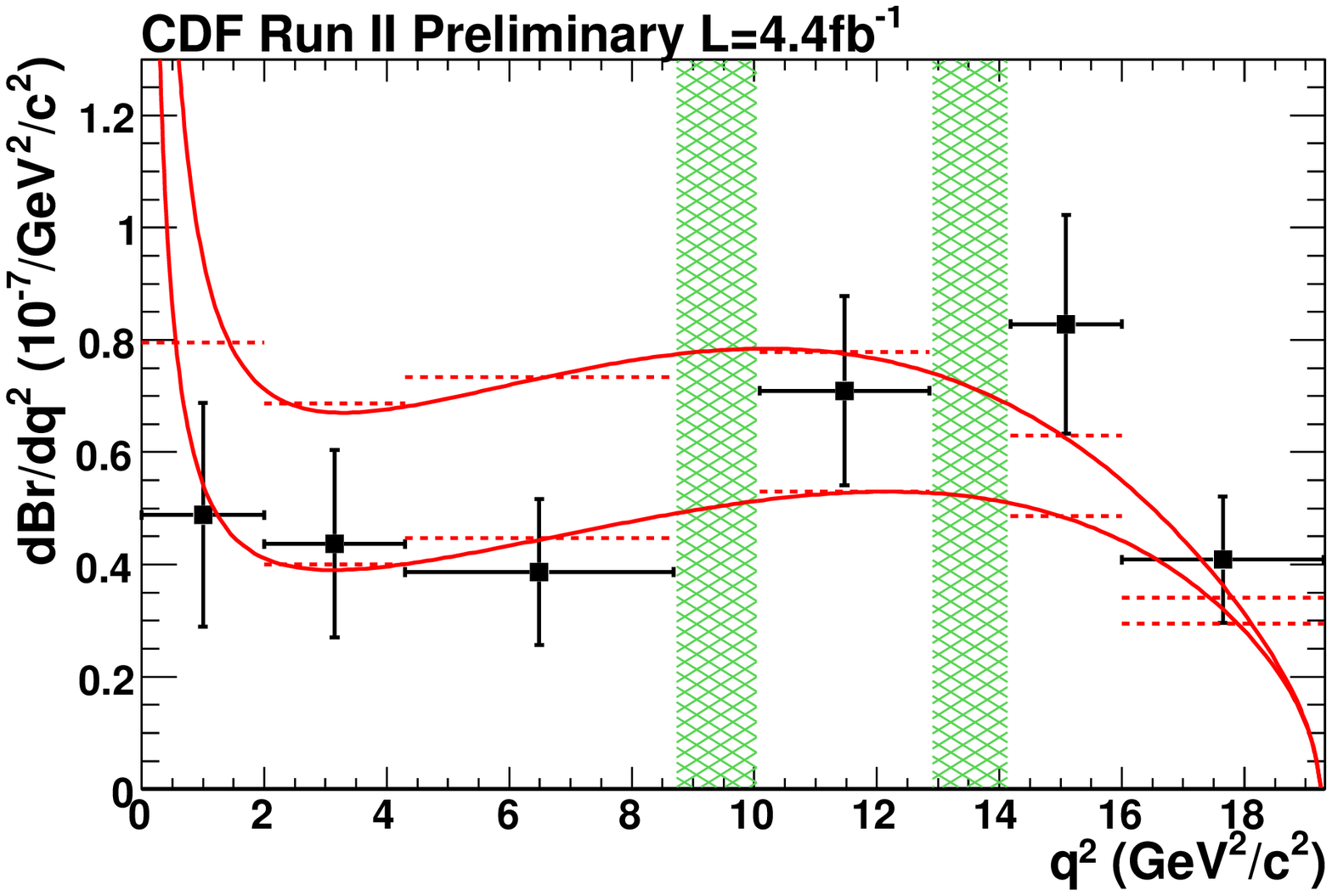}}
}
\caption{Differential branching ratio for $B^{+}\rightarrow K^{+}\mu^{+}\mu^{-}$ (left) and $B^{0}\rightarrow K^{*0}\mu^{+}\mu^{-}$ (right). The limits for the standard model expectation are shown in red.}
\label{diffbr}
\end{figure}

The differential branching ratio for the $K^{(*)}$ modes was measured in bins of $q^{2}=M_{\mu\mu}^{2}$, as shown in Figure~\ref{diffbr}. The limits on the standard model expectation are denoted by the red lines. The data points are consistent with these limits.

\begin{figure}[htbp]
\centerline{
\makebox{\includegraphics[width=0.33\textwidth]{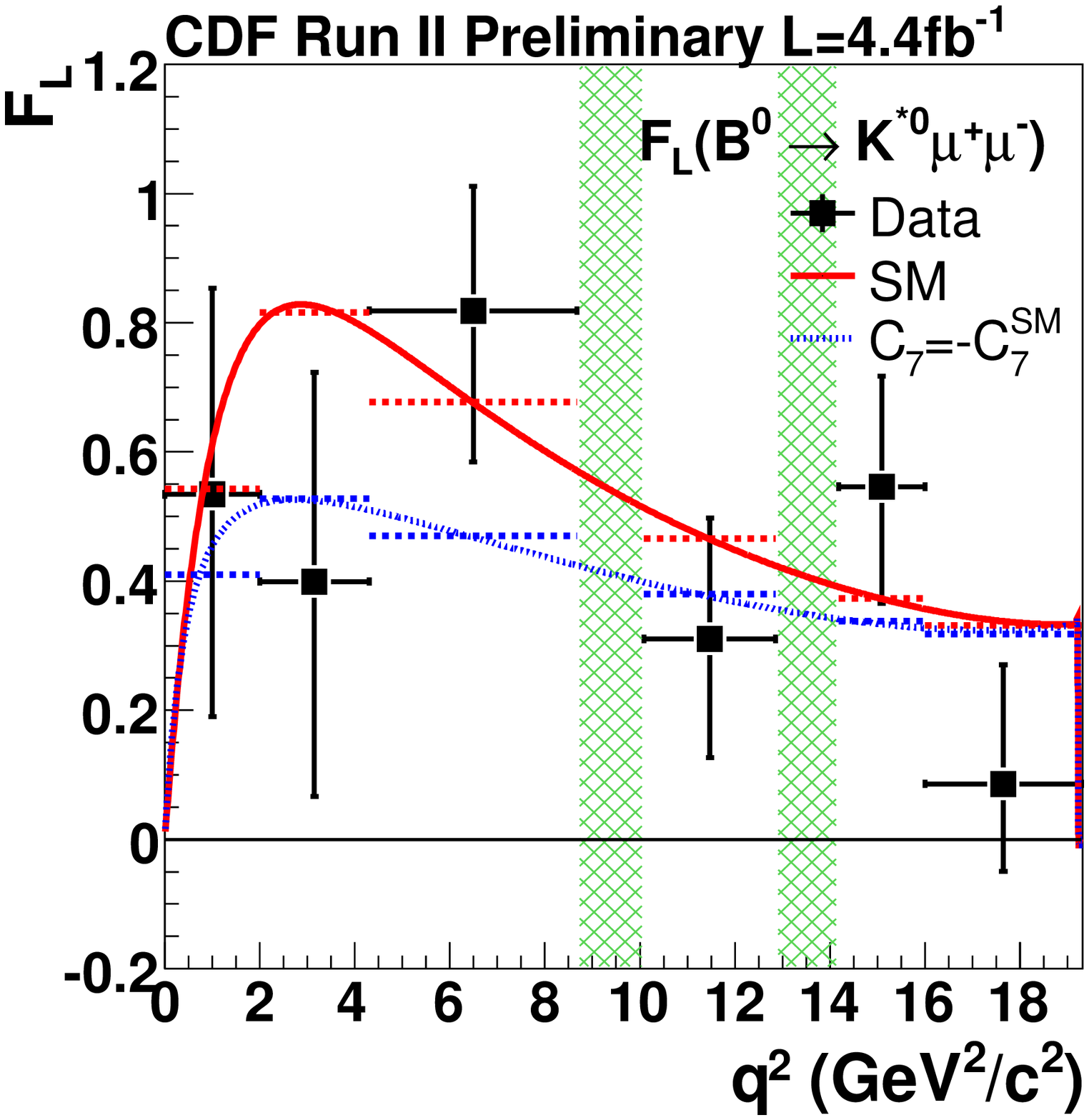}}
\makebox{\includegraphics[width=0.33\textwidth]{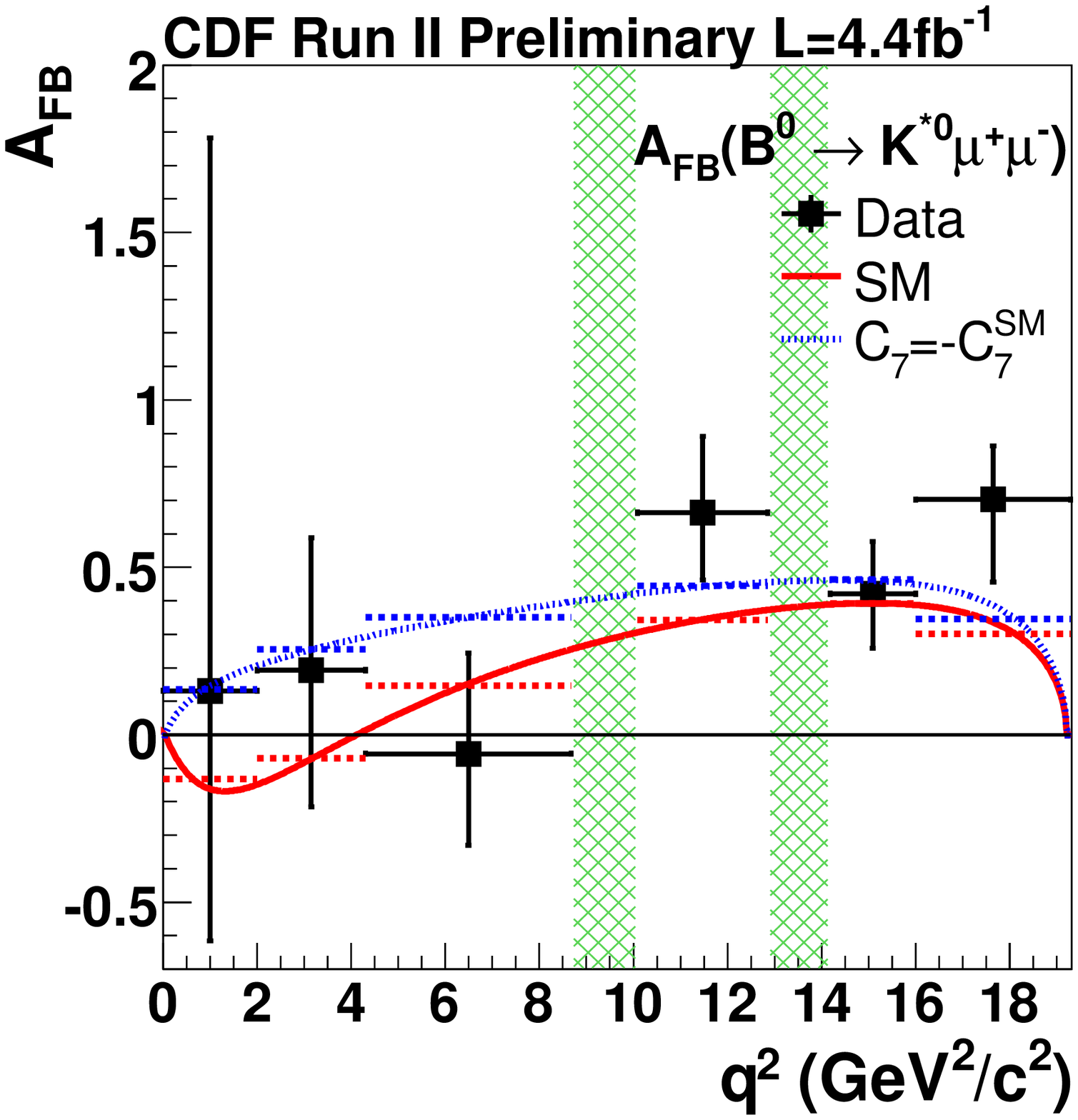}}
\makebox{\includegraphics[width=0.33\textwidth]{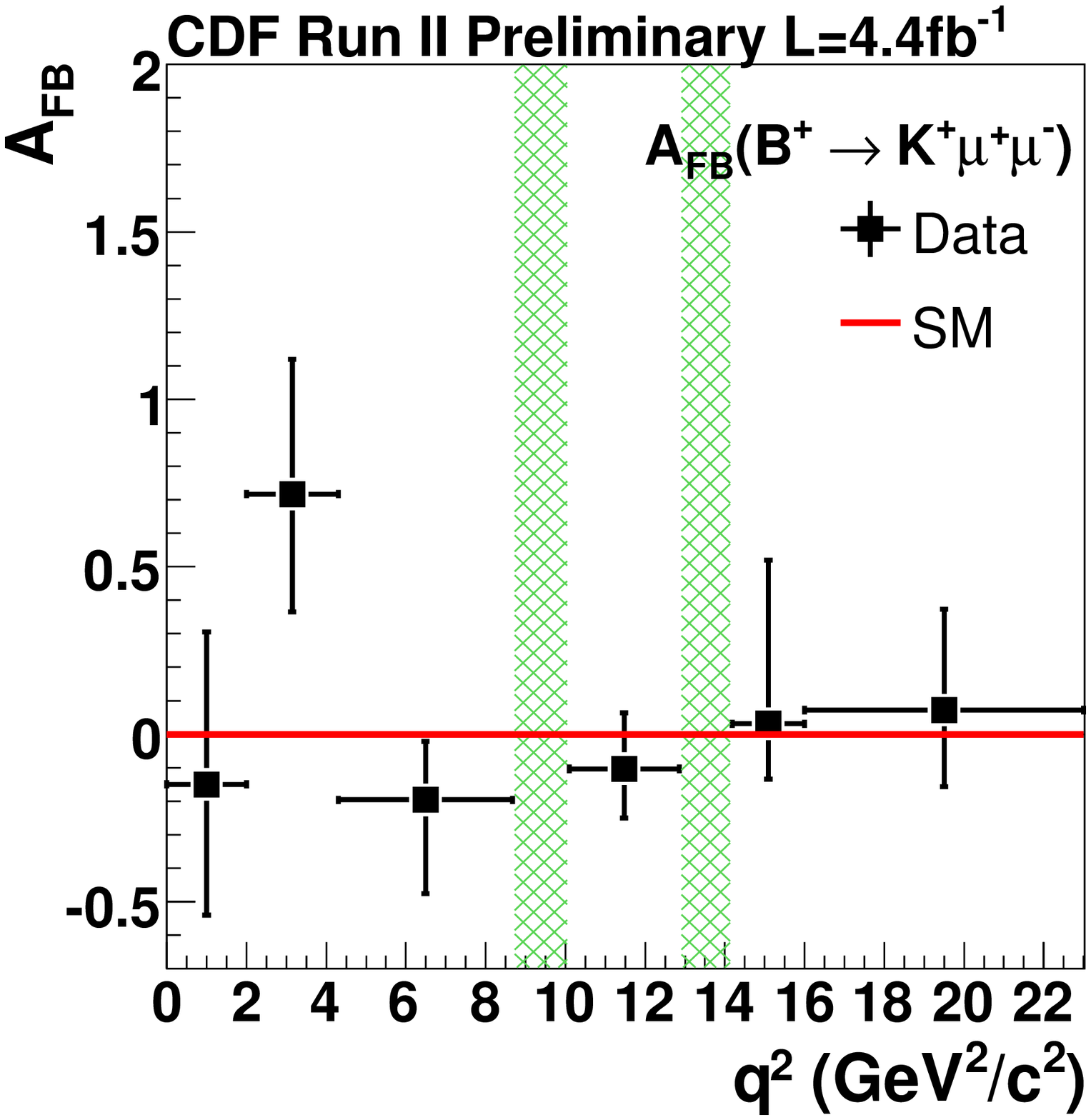}}
}
\caption{$K^{0*}$ polarization $F_{L}$ and $B^{0}\rightarrow K^{0*}\mu^{+}\mu^{-}$  and $B^{+}\rightarrow K^{+}\mu^{+}\mu^{-}$ forward-backward asymmetry $A_{FB}$.}
\label{AFB}
\end{figure}

The $K^{0*}$ polarization $F_{L}$ and $B^{0} \rightarrow K^{0*}\mu^{+}\mu^{-}$  and $B^{+}\rightarrow K^{+}\mu^{+}\mu^{-}$ forward-backward asymmetry $A_{FB}$ are also measured. An angular analysis is performed to extract $F_{L}$ and $A_{FB}$ in bins of $q^{2}=M_{\mu\mu}^{2}$. Results are shown in Fig.~\ref{AFB}. The standard model expectation is denoted by a red line, and a generic new physics scenario with flipped sign of the Wilson coefficient $C_{7}$ is indicated by a blue line. The precision of the measurement is not adequate to determine which scenario is favored. The results are consistent and competitive with $B$ factory measurements~\cite{aslnote}.

\section{Measurement of \textit{CP} violating phase $\boldsymbol{\sin2\beta_{s}}$ in $\boldsymbol{B^{0}_{s}\rightarrow J/\psi \phi}$ decays}

The $CP$ violating phase $\sin2\beta_{s}$ quantifies the $CP$ violation in the interference between the amplitudes from $B^{0}_{s}\rightarrow J/\psi \phi$ and $B^{0}_{s}\rightarrow \bar{B}^{0}_{s} \rightarrow J/\psi \phi$ decays. In the latter case, the $B^{0}_{s}$ mixes into its antiparticle before decaying, through the exchange of $W$ bosons and off-shell up type quarks in a mixing box diagram. The standard model expectation for $\sin2\beta_{s}$ is small, but new physics participation in the $B^{0}_{s}$ mixing box diagram could result in a large value for $\sin2\beta_{s}$.

The measurement is made on 2.8 fb$^{-1}$ of data collected using the di-muon trigger. Approximately 3000 $B^{0}_{s} \rightarrow J/\psi\phi \rightarrow \mu^{+}\mu^{-}K^{+}K^{-}$ events are reconstructed. An un-binned maximum likelihood fit to mass, lifetime, and final state angular distributions is used to extract $\sin2\beta_{s}$. The measurement's power is enhanced by using flavor tagging algorithms, which determine whether the candidate meson was a $B^{0}_{s}$ or a $\bar{B}^{0}_{s}$ at production. 

\begin{figure}[htbp]
\centerline{
\makebox{\includegraphics[width=0.3\textwidth]{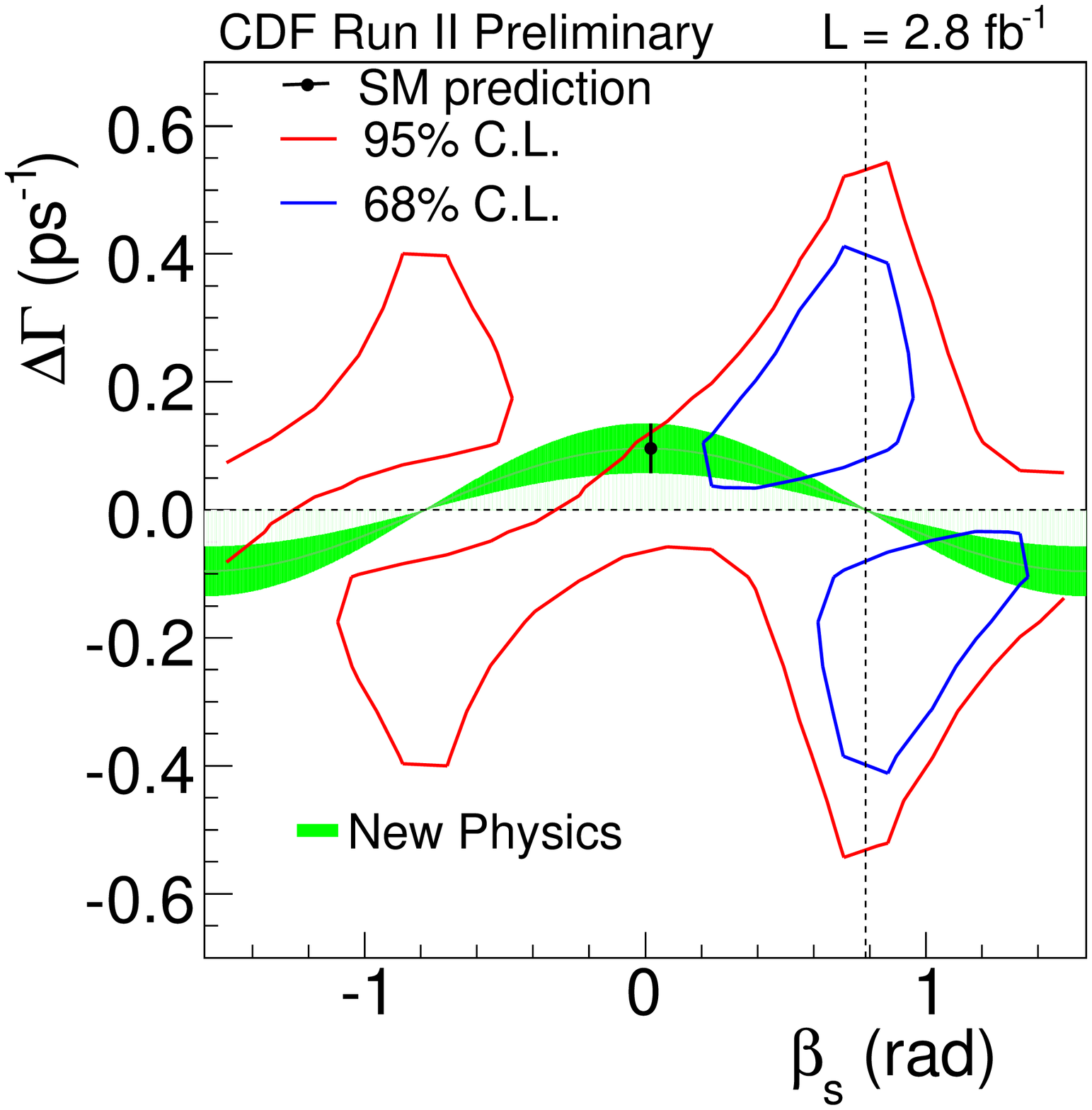}}
\makebox{\includegraphics[width=0.38\textwidth]{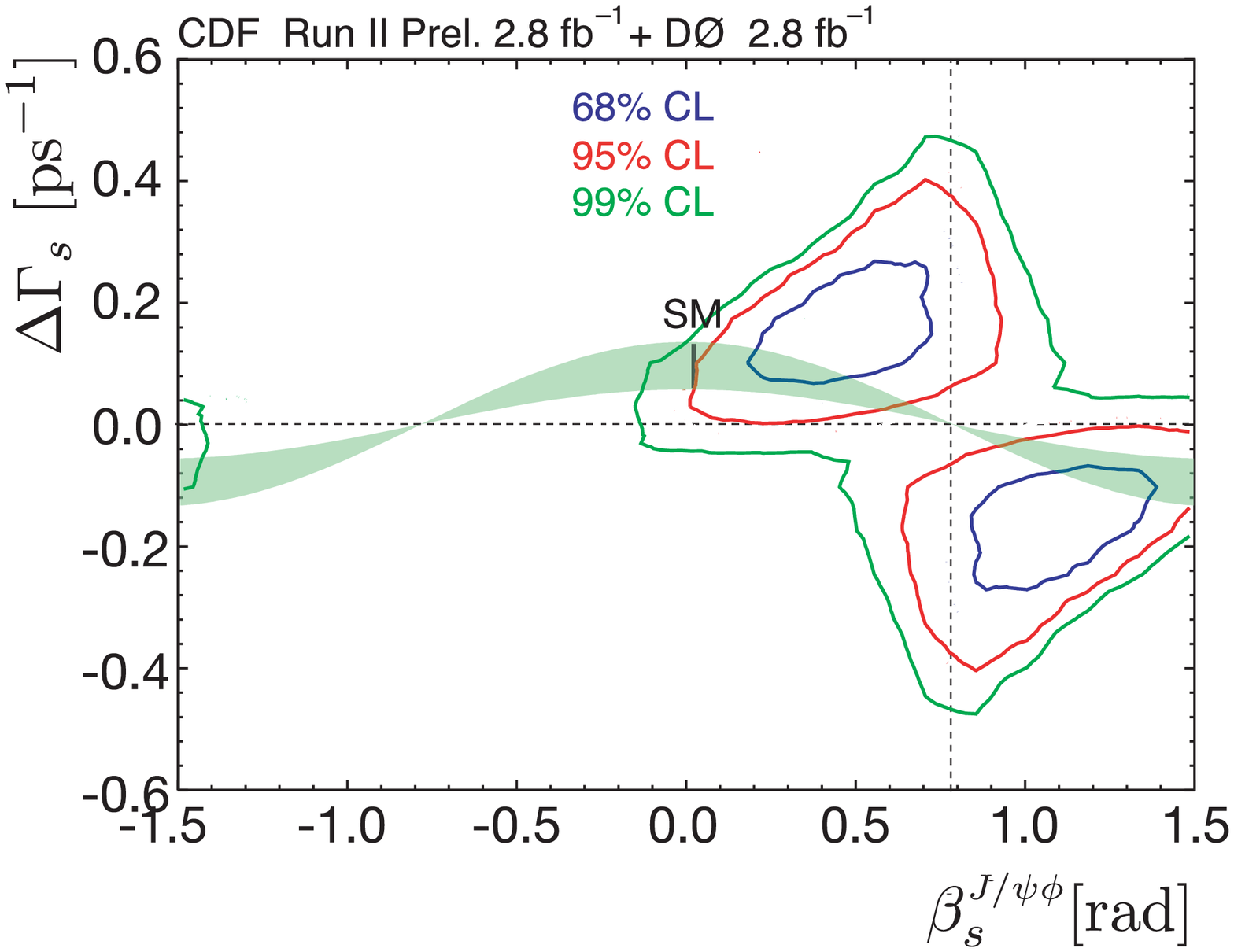}}
}
\caption{$\beta_{s}-\Delta\Gamma$ confidence region, CDF only (left), Tevatron combined result (right). The green bands show the allowed region assuming mixing-induced $CP$ violation.}
\label{bpdG}
\end{figure}

The left plot in Fig.~\ref{bpdG} shows a confidence region in $\beta_{s}$ and $\Delta\Gamma$ space, where $\Delta\Gamma$ is the decay width difference between the light and heavy $B^{0}_{s}$ mass eigenstates. The standard model expectation lies within the 95\% confidence level, with a p-value of 7\%. This means that the probability the true value is the standard model expectation and the observed data are a fluctuation is 7\%~\cite{sin2betasnote}.

The CDF measurement has been combined with an analogous measurement from the D\O~experiment. The result is shown in the right plot in Fig.~\ref{bpdG}. The standard model expectation falls outside the 95\% confidence level, indicating a 2$\sigma$ deviation of the contours from the SM prediction~\cite{combinationnote}. 

\section{$\boldsymbol{B \rightarrow J/\psi X}$ lifetimes}

The measurement of $B$ meson lifetimes is an important test of heavy quark effective theory. The $B^{+}$, $B^{0}$, and $\Lambda^{0}_{b}$ lifetimes are measured using 4.3 fb$^{-1}$ of data collected with the CDF di-muon trigger. For $B^{+}\rightarrow J/\psi K^{+}$, 45,000$\pm$230 events are reconstructed, for $B^{0}\rightarrow J/\psi K^{0*}$, 16,860$\pm$140 events, for $B^{0}\rightarrow J/\psi K^{0}_{s}$, 12,070$\pm$120 events, and for $\Lambda^{0}_{b}\rightarrow J/\psi\Lambda^{0}$, 1,710$\pm$50 events.

\begin{figure}[htbp]
\centerline{
\makebox{\includegraphics[width=0.35\textwidth]{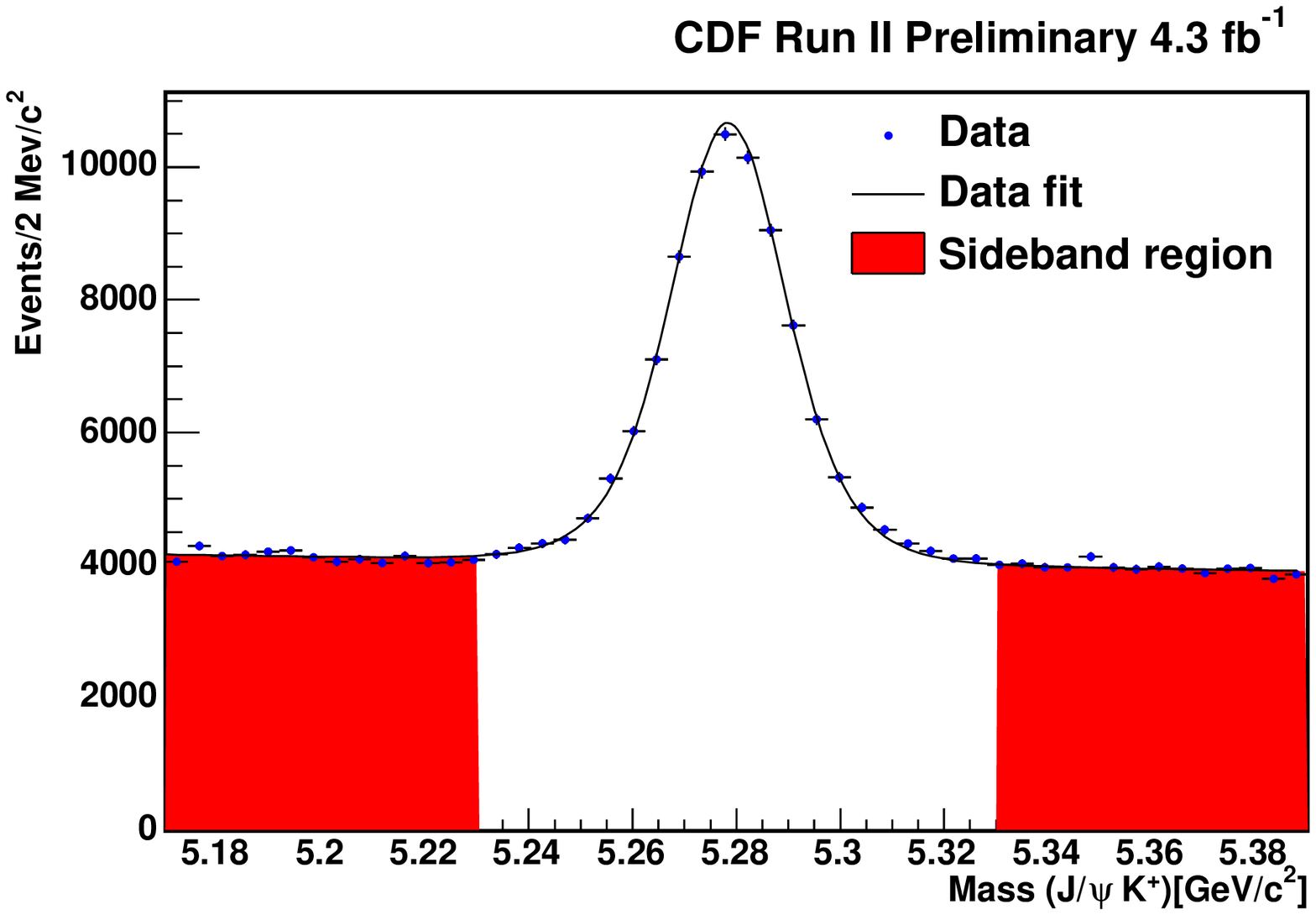}}
\makebox{\includegraphics[width=0.35\textwidth]{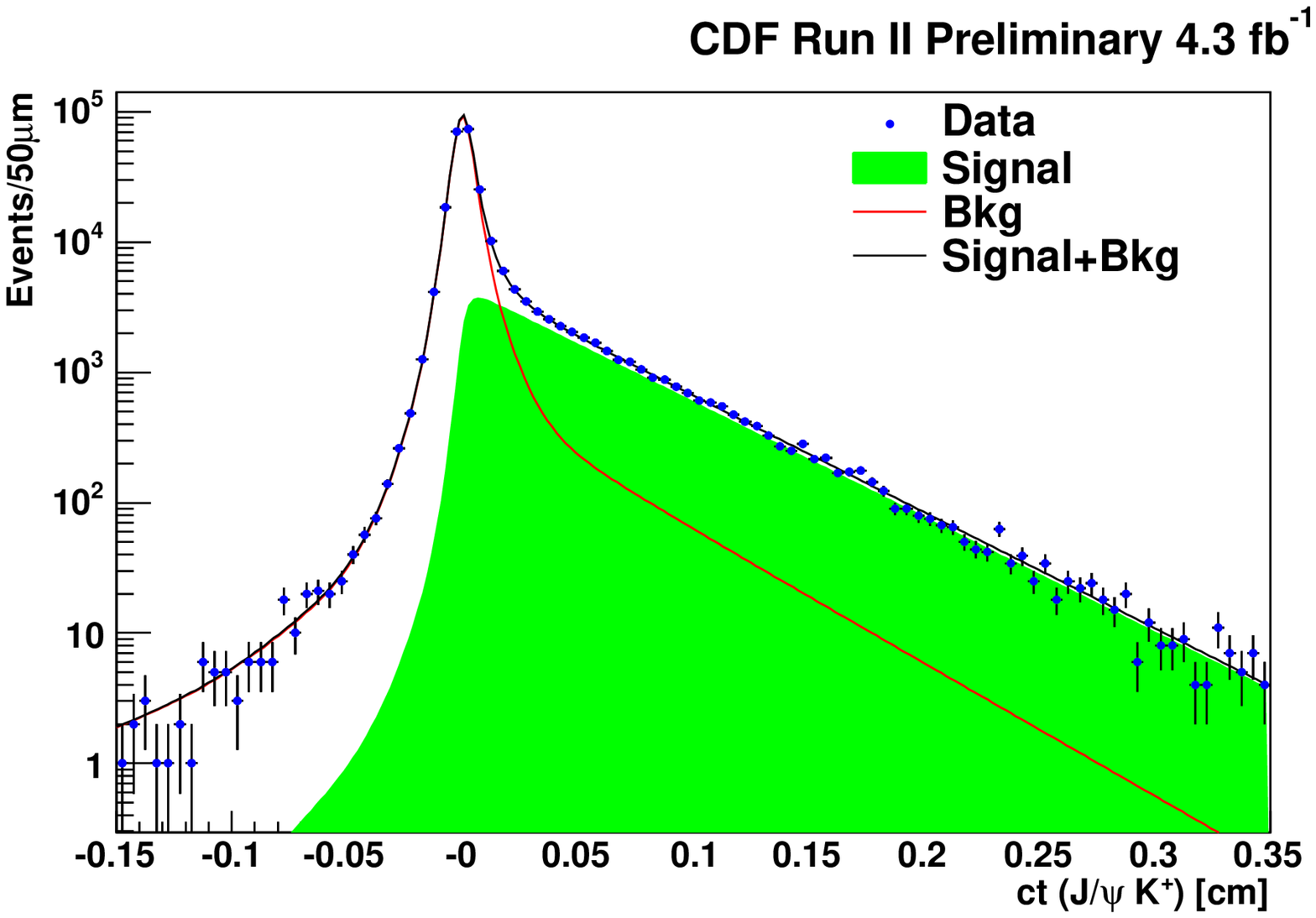}}
}
\caption{Fit projections for mass (left) and proper time (right).}
\label{lifetimefp}
\end{figure}

A combined fit to mass and lifetime is used to determine the lifetime. The fit projection for the mass is shown in the left-most plot in Fig.~\ref{lifetimefp}. The fit projections for proper time and its error in the signal region are shown in the center and right-most plots.

The lifetimes, which are the world's best measurements, are the following~\cite{jpslifetimenote}:

\begin{itemize}
\item{$\tau(B^{+})$ = 1.639 $\pm$ 0.009(stat) $\pm$ 0.009(syst) ps}
\item{$\tau(B^{0})$ = 1.507 $\pm$ 0.010(stat) $\pm$ 0.008(syst) ps}
\item{$\tau(\Lambda^{0}_{b})$ = 1.537 $\pm$ 0.045(stat) $\pm$ 0.014(syst) ps}
\item{$\tau(B^{+})/\tau(B^{0})$ = 1.088 $\pm$ 0.009(stat) $\pm$ 0.004(syst)}
\item{$\tau(\Lambda^{0}_{b})/\tau(B^{0})$ = 1.020 $\pm$ 0.030(stat) $\pm$ 0.008(syst).}
\end{itemize}

\section{Resonance structure of $\boldsymbol{\Lambda^{0}_{b}\rightarrow \Lambda^{+}_{c}\pi^{-}\pi^{+}\pi^{-}}$}

The $\Lambda^{0}_{b}$ meson can decay to an intermediate resonant states before decaying to the final state $\Lambda_{c}\pi^{-}\pi^{+}\pi^{-}$. This measurement is the first observation of these decay modes and their relative branching fractions.

The measurement is performed on 2.4 fb$^{-1}$ of data, collected using the two-track trigger. The following resonant modes are observed, with yields:

\begin{itemize}
\item{$\Lambda^{0}_{b}\rightarrow \Lambda_{c}(2595)^{+}\pi^{-} \rightarrow \Lambda^{+}_{c}\pi^{-}\pi^{+}\pi^{-}$, 46.6$\pm$9.7 events}
\item{$\Lambda^{0}_{b}\rightarrow \Lambda_{c}(2625)^{+}\pi^{-} \rightarrow \Lambda^{+}_{c}\pi^{-}\pi^{+}\pi^{-}$, 114$\pm$13 events}
\item{$\Lambda^{0}_{b}\rightarrow \Sigma_{c}(2455)^{++}\pi^{-}\pi^{-} \rightarrow \Lambda^{+}_{c}\pi^{-}\pi^{+}\pi^{-}$, 81$\pm$15 events}
\item{$\Lambda^{0}_{b}\rightarrow \Sigma_{c}(2455)^{0}\pi^{+}\pi^{-} \rightarrow \Lambda^{+}_{c}\pi^{-}\pi^{+}\pi^{-}$, 41.5$\pm$9.3 events}
\item{$\Lambda^{0}_{b}\rightarrow \Lambda^{+}_{c}\rho^{0}\pi^{-} + \Lambda^{+}_{c}3\pi(other) \rightarrow \Lambda^{+}_{c}\pi^{-}\pi^{+}\pi^{-}$, 610$\pm$88 events.}
\end{itemize}

The relative branching fractions are as follows~\cite{lambdabnote}:

\begin{itemize}
\item{$\frac{BR(\Lambda^{0}_{b}\rightarrow \Lambda_{c}(2595)^{+}\pi^{-} \rightarrow \Lambda^{+}_{c}\pi^{-}\pi^{+}\pi^{-})}{BR(\Lambda^{0}_{b}\rightarrow \Lambda^{+}_{c}\pi^{-}\pi^{+}\pi^{-} (all))}$ = 2.5$\pm$0.6(stat)$\pm$0.5(syst)$\times$ 10$^{-2}$}
\item{$\frac{BR(\Lambda^{0}_{b}\rightarrow \Lambda_{c}(2625)^{+}\pi^{-} \rightarrow \Lambda^{+}_{c}\pi^{-}\pi^{+}\pi^{-})}{BR(\Lambda^{0}_{b}\rightarrow \Lambda^{+}_{c}\pi^{-}\pi^{+}\pi^{-} (all))}$ = 6.2$\pm$1.0(stat)$^{+1.0}_{-0.9}$(syst)$\times$ 10$^{-2}$}
\item{$\frac{BR(\Lambda^{0}_{b}\rightarrow \Sigma_{c}(2455)^{++}\pi^{-}\pi^{-} \rightarrow \Lambda^{+}_{c}\pi^{-}\pi^{+}\pi^{-})}{BR(\Lambda^{0}_{b}\rightarrow \Lambda^{+}_{c}\pi^{-}\pi^{+}\pi^{-} (all))}$ = 5.2$\pm$1.1(stat)$\pm$0.8(syst)$\times$ 10$^{-2}$}
\item{$\frac{BR(\Lambda^{0}_{b}\rightarrow \Sigma_{c}(2455)^{0}\pi^{+}\pi^{-} \rightarrow \Lambda^{+}_{c}\pi^{-}\pi^{+}\pi^{-})}{BR(\Lambda^{0}_{b}\rightarrow \Lambda^{+}_{c}\pi^{-}\pi^{+}\pi^{-} (all))}$ = 8.9$\pm$2.1(stat)$^{+1.2}_{-1.0}$(syst)$\times$ 10$^{-2}$}
\end{itemize}

\section{$\boldsymbol{\Upsilon(1S)}$ Polarization}

Measurements of the $J/\psi$ and $\Upsilon(1S)$ polarization are used to test the predictions of non-relativistic QCD. The $J/\psi$ polarization measurement disagrees with theory, making $\Upsilon(1S)$ polarization the subject of much interest.

\begin{figure}[htbp]
\centerline{
\makebox{\includegraphics[width=0.27\textwidth]{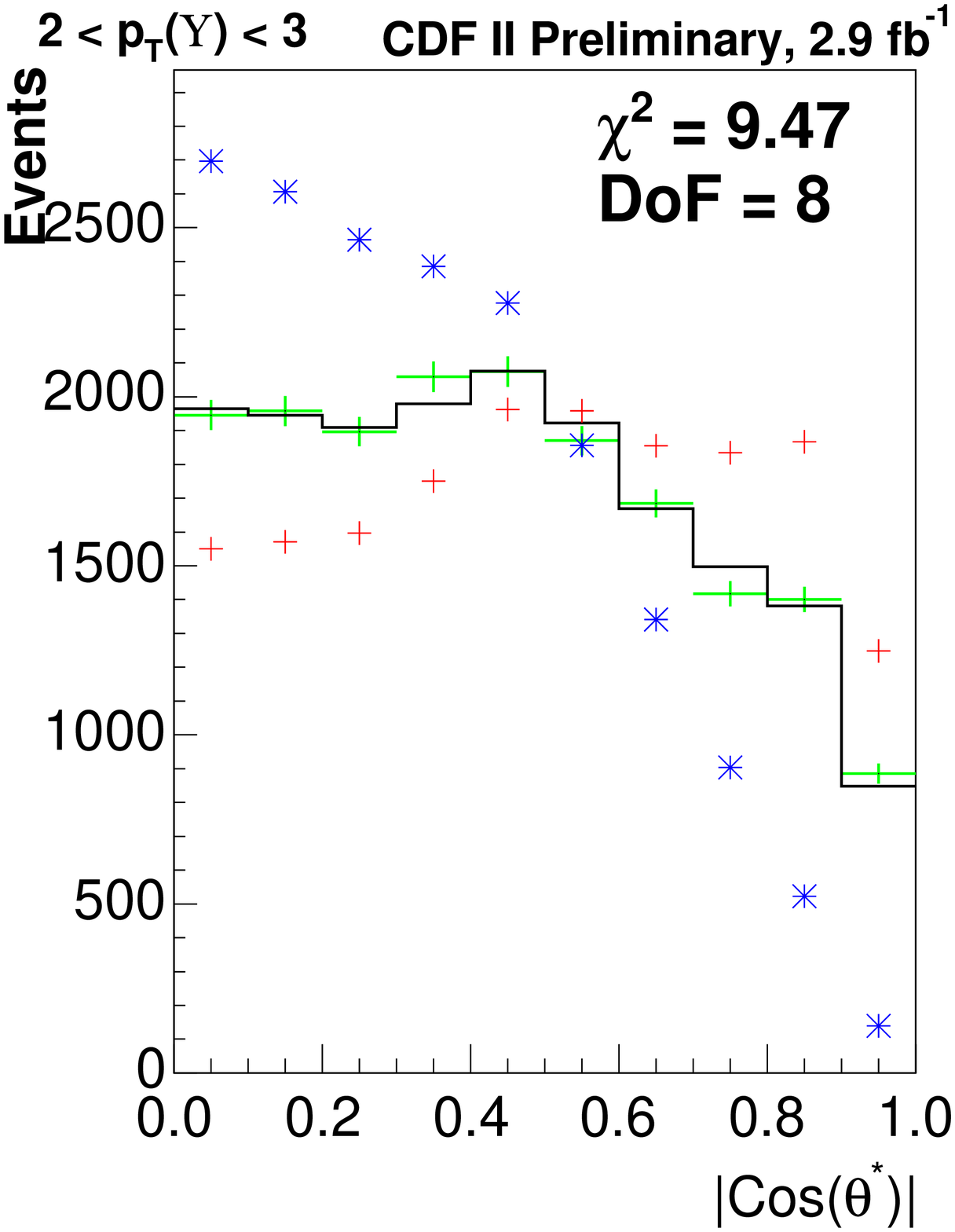}}
\makebox{\includegraphics[width=0.52\textwidth]{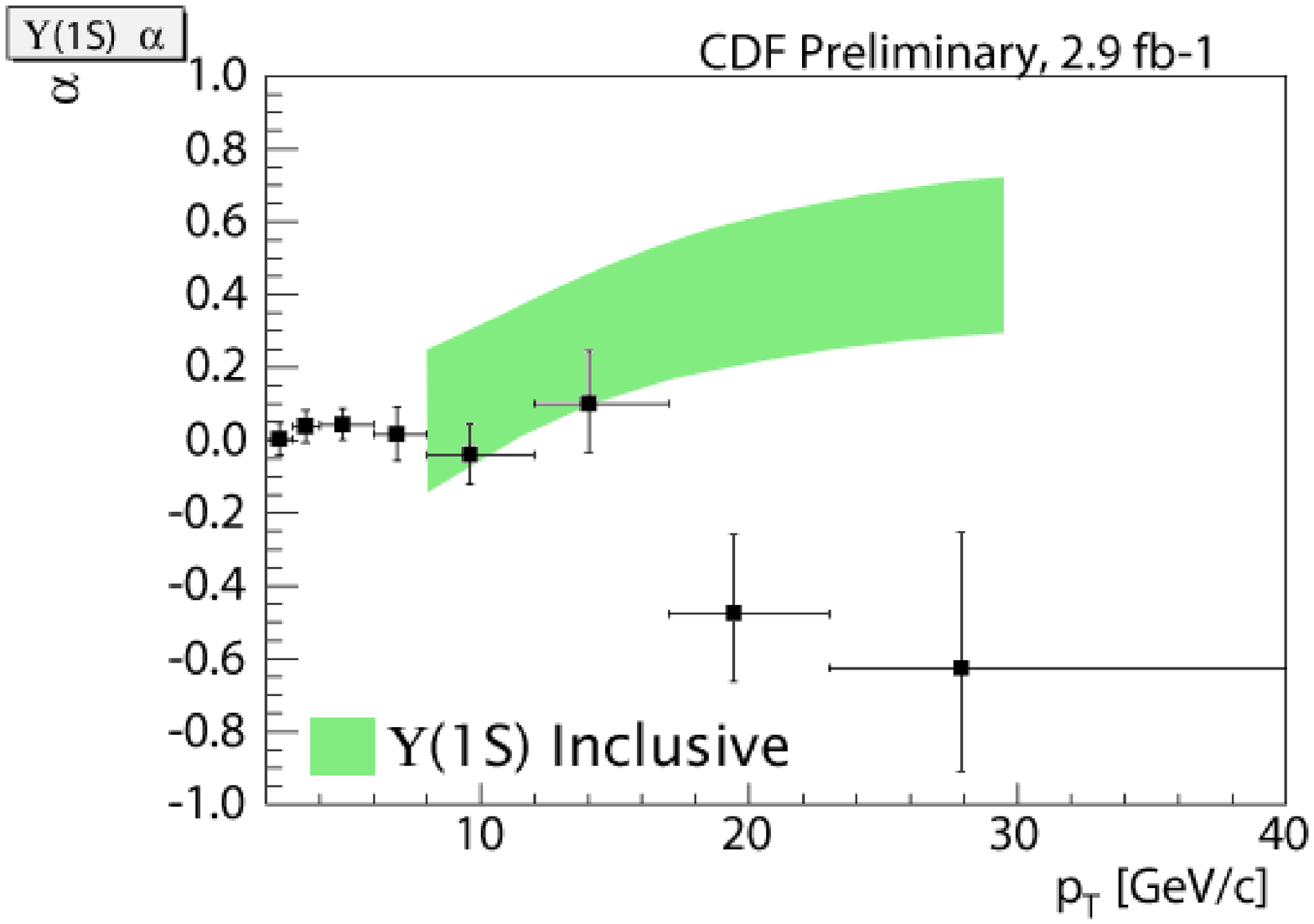}}
}
\caption{Polar angle distribution for $\mu^{+}$ with transverse (red) and longitudinal (blue) Monte Carlo templates (left). Polarization as a function of $p_{T}$, with NRQCD predictions (green).}
\label{upsipol}
\end{figure}

The measurement is made on 2.9 fb$^{-1}$ of data, collected with a di-muon trigger. The polarization parameter $\alpha$ is determined by studying the distribution of the angle $|\cos(\theta^{*})|$ associated with the positive muon from $\Upsilon(1S) \rightarrow \mu^{+} \mu^{-}$. By comparing the observed angular distributions in different $p_{T}(\Upsilon)$ bins with templates of fully transverse and longitudinal Monte Carlo (as shown in the left plot in Fig.~\ref{upsipol}), the polarization can be determined. The polarization parameter $\alpha$ is shown as a function of $p_{T}$ in the right plot in Fig.~\ref{upsipol}, with the NRQCD prediction in green. The data are in poor agreement with theory at high $p_{T}$~\cite{upsinote}.

\end{document}